\documentclass[USenglish,twocolumn]{article}

\usepackage[utf8]{inputenc}
\usepackage[big]{dgruyter_NEW}

\usepackage[mathscr]{euscript}
\usepackage{textcomp}

\usepackage[]{amsmath}
\usepackage{amssymb,graphicx}
\usepackage[]{graphicx}

\begin{document}

\articletype{Research Article{\hfill}Open Access}
\author[1]{Pengchao Yu}
\author*[2]{Volodymyr~I.~Fesenko}
\author[3]{Vladimir~R.~Tuz}

  \affil[1]{International Center of Future Science, College of Physics, Jilin University, 2699  Qianjin Str., Changchun 130012, China, E-mail: yupc17@mails.jlu.edu.cn}

  \affil[2]{International Center of Future Science, Jilin University, 2699  Qianjin Str., Changchun 130012, China; Department of Microwave Electronics, Institute of Radio Astronomy of National Academy of Sciences of Ukraine, 4, Mystetstv Street, Kharkiv 61002, Ukraine, E-mail: volodymyr.i.fesenko@gmail.com}

  \affil[3]{International Center of Future Science, State Key Laboratory on Integrated Optoelectronics, College of Electronic Science and Engineering, Jilin University, 2699  Qianjin Str., Changchun 130012, China; Department of Theoretical Radio Physics, Institute of Radio Astronomy of National Academy of Sciences of Ukraine, 4, Mystetstv Street, Kharkiv 61002, Ukraine, E-mail: tvr@jlu.edu.cn; tvr@rian.kharkov.ua}

  \title{\huge Dispersion features of complex waves in a~graphene-coated semiconductor nanowire}

  \runningtitle{Dispersion features of complex waves in a~graphene-coated semiconductor nanowire}


  \begin{abstract}
{Dispersion features of a graphene-coated semiconductor nanowire operating in the terahertz frequency band are consistently studied in the framework of a special theory of complex waves. Detailed classification of the waveguide modes was carried out based on the analysis of characteristics of the phase and attenuation constants obtained from the complex roots of  characteristic equation. With such a treatment, the waves are attributed to the group of either `proper' or `improper' waves, wherein their type is determined as the trapped surface waves, fast and slow leaky waves, and surface plasmons. The dispersion curves of axially symmetric TM$_{0n}$ and TE$_{0n}$ modes, as well as non-symmetric hybrid EH$_{1n}$ and HE$_{1n}$ modes were plotted and analyzed in details, and both radiative regime of leaky waves and guided regime of trapped surface waves are identified. Peculiarities of propagation of the TM modes of surface plasmons were revealed. Two sub-regions of existence of surface plasmons were found out where they appear as propagating and reactive waves. The cut-off conditions for higher order modes were correctly determined.}
\end{abstract}
  \journalname{Nanophotonics}
  \startpage{1}
\maketitle
\section{Introduction}
\label{sec:intro}
Rapid progress in nanotechnologies allows producing advanced highly integrated devices capable to operate in the terahertz range of electromagnetic radiation. In such  devices, nanowires, and especially semiconductor nanowires, appear as a main constitutive element possessing both photonic and plasmonic functionality and demonstrating unique optical, magnetic, thermal, electronic and mechanical characteristics \cite{Wei_NP_2012, Dasgupta_ADMA_2014, Kuzmin_NP_2018}. The modern growth techniques allow to synthesize semiconductor nanowires having desired compositions, heterojunctions and architectures. In particular, they have been already incorporated into diverse integrated nanosystems designed for performing computational and communication operations, biological and chemical sensory procedures as well as for energy conversion and storage (see, for instance, \cite{Sirbuly_PNAS_2005, Yan_NatPh_2009, Wei_NP_2012, Dasgupta_ADMA_2014, Abujetas_ACS_2015, Zhou_OE_2017, Kuzmin_NP_2018} and references therein).

Semiconductor nanowires belong to a class of optical cylindrical waveguides whose core typically has a diameter smaller than a micrometer. Due to such  small diameter of the core, in the terahertz range nanowires usually operate under the subwavelength propagation conditions which differ from those of the conventional optical waveguides whose diameter is larger than the wavelength. According to the electromagnetic theory, subwavelength modes existing in the nanowires are generally complex waves being either surface (propagating) or leaky (radiative) ones whose propagation constant is a complex quantity even in the absence of losses in the nanowire constitutive materials (i.e., the complex waves are associated with the modes whose propagation constant is a complex quantity, rather than purely real or purely imaginary one \cite{Tamir_IEEE_1963, Veselov_book_1988, Ishimaru_book_1991, Marcuse_book_1991}).
  
Comparing with ordinary large-diameter optical waveguides, subwavelength-diameter nanowires demonstrate enhanced evanescent fields, tight light confinement and large waveguide dispersions \cite{Tong_OE_2004}. Remarkably, under the subwavelength propagation conditions the nanowires carry the main fraction of the modal intensity outside the core, and this effect is a basis of many important practical applications. In particular, such a modal energy distribution allows to obtain significant level of interaction between the waveguide modes and adsorbed molecules located outside the waveguide in biochemical sensors \cite{Zhang_2015,Sirbuly_PNAS_2005}. It is also used in photo-voltaic systems to increase light concentration, as well as to tune and enhance fundamental absorption properties and power conversion efficiency \cite{Cao_NM_2009, Zhou_OE_2017}. Besides, in the radiative regime nanowires operate as  efficient low-loss omnidirectional leaky-wave antennas \cite{Kim_2003}.

In addition to the surface and leaky waves, the TM modes of surface plasmons are also supported by the semiconductor nanowires below the plasma frequency of their core material. Thus, the metallic character of doped semiconductors makes it possible to excite surface plasmons at terahertz. As the carrier densities in semiconductors are much lower than those in metals, the plasma frequency is much smaller, being typically at mid- or far-infrared range. Therefore, the permittivity of semiconductors at terahertz frequencies is comparable to that of metals at optical frequencies. A decisive advantage of semiconductors is that their carrier density and mobility, and consequently the surface plasmons, can be easily controlled by thermal excitation of free carriers \cite{Gomez_ApplPhysLett_2006}.

In optics, in order to achieve better confinement of surface plasmons, the nanowires typically are coated by noble metals. Unfortunately, at terahertz frequencies noble metals demonstrate severe Ohmic losses and therefore they are not always suitable for plasmons guiding at this range \cite{Schroter_PhysRevB_2001}. For today there is a confidence that at terahertz frequencies the most suitable material to cover the nanowires is graphene \cite{Jablan_PhysRevB_2009}. Compared to surface plasmons existed on noble metals, graphene provides lower Ohmic losses, tight subwavelength confinement, and can support plasmons of both TM and TE polarizations \cite{Mikhailov_PRL_2007, Kuzmin_SR_2016}. Moreover, a graphene sheet can tightly coat the semiconductor nanowire due to van der Waals force \cite{Wu_OL_2014}, and then the possibility arises to tune the carrier density of graphene by changing either chemical potential or by using electrostatic bias field. It opens prospects for numerous practical applications of graphene-coated semiconductor nanowires \cite{Jablan_PhysRevB_2009, Gao_OE_2014, Zhu_JOSAB_2015, Correas-Serrano_IEEE_2015,Fuscaldo_2015, Cuevas_2016, Kuzmin_SR_2016,Gao_SR_2016}. 

The dispersion features of surface (guided) waves as well as surface plasmons propagating through a graphene-coated nanowire are usually studied within the framework of the classical electromagnetic theory \cite{Gao_OE_2014} developed for dielectric waveguides \cite{Marcuse_book_1991,Unger_book_1977}. In particular, within this theory, the waveguide modes of  both surface waves and surface plasmons are classified, the number of supported modes and single-mode condition are obtained, and a formula for the modal cut-offs calculation is derived. The propagation length of the TM modes of surface plasmons on a graphene sheet coating the nanowire is estimated \cite{Gao_OptLett_2014}. Propagation conditions of the weakly localized TE modes of surface plasmons in a cylindrical graphene-based waveguide were studied \cite{Kuzmin_SR_2016}. The tunable performance of a graphene-coated nanowire by varying the graphene Fermi level is also analyzed.

Nevertheless, the theory of wave propagation in graphene-coated semiconductor nanowires is far from being completed. In general, the modes supported by such nanowires a complex waves \cite{Fuscaldo_2015, Cuevas_2016} which possess different nature being evanescent propagating or radiative ones. It greatly complicates their identification and analysis, and requires application of a special theory. In this paper we perform our study of dispersion features of a graphene-coated semiconductor nanowire involving the theory of complex waves \cite{Tamir_IEEE_1963, Ishimaru_book_1991, Veselov_book_1988}. It allows us to properly classify the modes, reveal the conditions for the mode transition between different wave types, and accurately determine their cut-off frequencies.

\section{Outline of Problem} 
\label{sec:problem}
\subsection{Description of a Graphene-Coated Semiconductor Nanowire}
\begin{figure}[!ht]
\centering
\includegraphics[width=0.9\linewidth]{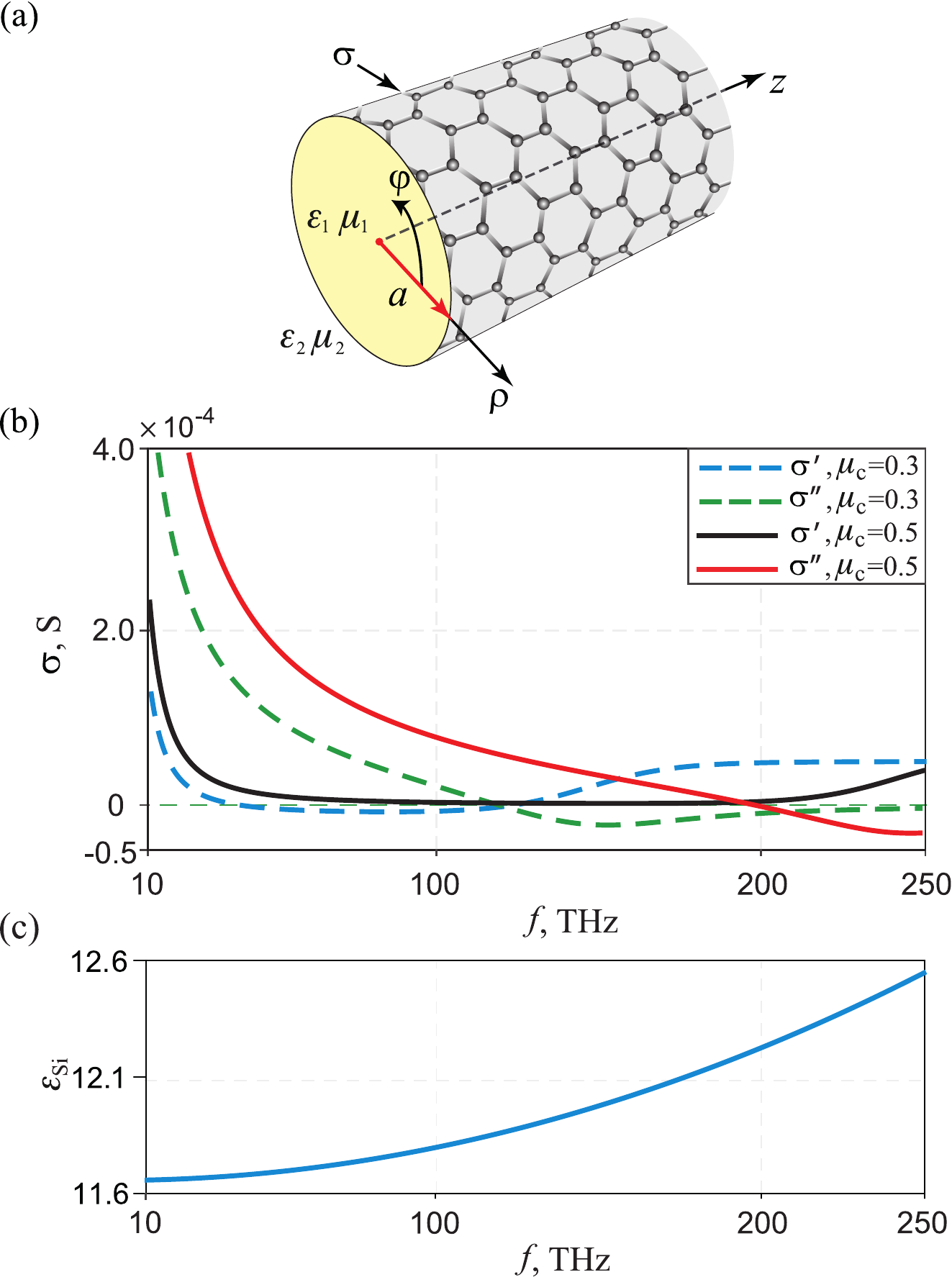}
\caption{(a) Schematic of a graphene-coated semiconductor nanowire and dispersion characteristics of (b) graphene conductivity $\sigma$ for different values of the chemical potential $\mu_c$, and (c) silicon permittivity $\varepsilon_{\textrm{Si}}$ at fixed temperature $T=300$~K.}
\label{fig:fig_1}
\end{figure}

In this paper it is our goal to classify modes and study dispersion features of complex waves of a cylindrical waveguide made of a semiconductor nanowire coated by a graphene sheet (Fig.~\ref{fig:fig_1}a). The semiconductor core has the radius $a$ and permittivity $\varepsilon_1$. The whole waveguide is surrounded by a medium with permittivity $\varepsilon_2$. Permeabilities of the core and surrounding medium are $\mu_1$ and $\mu_2$, respectively. Due to the azimuthal symmetry of the waveguide under study we use a cylindrical polar coordinate system ($\rho$, $\varphi$, $z$) assuming the symmetry axis of nanowire coincides with the $z$-axis of the coordinate system.

We suppose the nanowire operating in the terahertz band ($10-250$~THz). Ignoring the quantum finite-size effect of graphene, the nanowire coating is further treated as an infinitely thin graphene sheet ($t_g\to 0$) having the macroscopic surface conductivity $\sigma$ dependent on the angular frequency $\omega=2\pi f$, chemical potential $\mu_c$, ambient temperature $T$, and charge carries scattering rate $\Gamma$. The surface conductivity of graphene consists of intraband and interband contributions $\sigma=\sigma_{intra}+\sigma_{inter}$, which are described by Kubo formalism \cite{Falkovsky_PhysRev_2007} 
\begin{equation}
\begin{split}
& \sigma_{intra} = \frac{2 i e^2k_BT}{\hslash^2 \pi \left(\omega+i \Gamma\right)}\left[2\cosh\left(\frac{\mu_c}{2k_BT} \right) \right],\\
&\sigma_{inter} = \frac{e^2}{4\hslash\pi}
\left[\frac{\pi}{2}+\arctan\left(\frac{\hslash\omega-2\mu_c}{2k_BT}\right) -\right.\\
&-\left.\frac{i}{2}\ln\frac{\left(\hslash\omega+2\mu_c\right)^2}{\left(\hslash\omega-2\mu_c\right)^2+\left(2k_BT\right)^2} \right].
\end{split}
\label{eq:Kubo} 
\end{equation}
Here $k_B$ is the Boltzmann constant, $\hslash$ is the reduced Planck constant, and $e$ is the electron charge. The chemical potential $\mu_c$  is related to the carriers density $N_c$ as $\mu_c=\hslash v_F \sqrt{\pi N_c}$, where $v_F\simeq 10^6$~m/s is the Fermi velocity of electrons in graphene. For clarity, the dispersion characteristics of the real $\sigma'$ and imaginary $\sigma''$ parts of the graphene conductivity in the frequency band of interest are plotted in Fig.~\ref{fig:fig_1}b at a fixed temperature $T$ and for different values of the chemical potential $\mu_c$. 

We should note, in the case when the thickness of graphene sheet $t_g$ is comparable to the waveguide radius $a$ in value, the sheet should be treated as a finite layer having permittivity 
\begin{equation}
\varepsilon_g = 1+\frac{i\sigma}{\varepsilon_0\omega t_g},
\label{eq:effective} 
\end{equation}
where $\varepsilon_0$ is the vacuum permittivity. 

The dispersion characteristic of permittivity of silicon forming the nanowire $\varepsilon_1=\varepsilon_0 \varepsilon_{\textrm{Si}}(\omega,T)$ can be approximated by the second order polynomial \cite{Li_JPCRD_1980} 
\begin{equation}
\varepsilon_{\textrm{Si}}(\omega,T)=\varepsilon_{\textrm{Si}}(T)+\frac{\omega^2L(T)}{4\pi^2 c^2}\left( A_0+A_1 T+A_2 T^2\right),
\label{eq:epsSi} 
\end{equation}
whose coefficients are generated from the experimental data involving fitting procedures. The extracted data cover several temperature ranges and frequency bands. In the frequency band of interest corresponding coefficients are: $A_0 = 0.8948$, $A_1 = 4.3977\times 10^{-4}$, $A_2 = 7.3835\times 10^{-8}$. The temperature function of permittivity is $\varepsilon_{\textrm{Si}}(T)=11.4445 + 2.7739\times 10^{-4} T + 1.705\times 10^{-6}T^2 - 8.1347\times 10^{-10}T^3$, while coefficients of the tuning function $L(T) = \exp[-3\Delta L(T)/L_{293}]$ should be chosen for the corresponding temperature range:  $20<T<293$~K, $\Delta L(T)/L_{293}=-0.021 - 4.149\times 10^{-7}T - 4.620\times 10^{-10}T^2 + 1.482\times 10^{-11}T^3$, and $293<T<750$~K, $\Delta L(T)/L_{293}=-0.071 + 1.887\times 10^{-6}T + 1.934\times 10^{-9}T^2 - 4.544\times 10^{-13}T^3$. A typical dispersion curve of permittivity of silicon calculated according to Eq.~(\ref{eq:epsSi}) at a fixed temperature $T$ is given in Fig.~\ref{fig:fig_1}c.

\subsection{Dispersion Relations}
Considering electromagnetic waves propagating through the waveguide along the $z$-axis direction, the electric and magnetic field vectors are expressed as
\begin{equation}
\vec{A}(\rho,\varphi,z,t) = \vec{A}(\rho)\exp\left[i(m\varphi+\beta z-\omega t)\right],
\label{eq:fields} 
\end{equation}
where $\vec{A}(\rho)$ is a cylindrical amplitude substituted for either magnetic $\vec{H}$ or electric $\vec{E}$ field strength, $m$ is the integer value known as the azimuthal wavenumber, and $\beta$ is the complex (longitudinal) propagation constant.

In this paper we follow the method described in \cite{Veselov_book_1988}, where a detailed solution procedure is given for the boundary-value problem regarding the modes of a cylindrical dielectric waveguide coated by a resistive film. This procedure implies deriving the electromagnetic field components inside ($\rho < a$, subscript `1') and outside ($\rho > a$, subscript `2') the dielectric waveguide, and imposing the boundary conditions on the tangential field components at the waveguide's circular wall coated by a resistive film ($\rho = a$). In our case, a graphene sheet acts as the resistive film with the surface conductivity $\sigma$, and the boundary conditions are written as follows: $E_{z1}=E_{z2}$, $E_{\varphi 1}=E_{\varphi 2}$, $H_{z2}-H_{z1}=-\sigma E_{\varphi 1}$, and $H_{\varphi 2}-H_{\varphi 1}=\sigma E_{z1}$. From these conditions a $4\times 4$ coefficient matrix is then constructed and the characteristic equation is derived under the assumption that the determinant of this matrix is zero to avoid nontrivial solutions. 

Generally the graphene-coated nanowire under study supports a set of the non-symmetric hybrid EH$_{mn}$ and HE$_{mn}$ modes, whose dispersion equation is expressed as
\begin{equation}
\begin{split}
& \omega^2P_1P_2- \frac{m^2 \beta^2}{a^2} \left(1-\frac{\kappa_1^2}{\kappa_2^2} \right)^2+ i\omega\sigma\ \frac{m^2 \beta^2}{a^2 \kappa_2^2}  \times \\
& \times \left[\mu_1\frac{\kappa_1^2}{\kappa_2^2}F(\kappa_1 a)-\mu_2 Q(\kappa_2 a)\right]-\mu_1\mu_2\sigma\frac{\omega^2}{\kappa_2^2} \times \\
& \times F(\kappa_1 a)Q(\kappa_2 a)\left(\sigma\kappa_1^2-i\omega P_1\right) +i\omega\sigma\kappa_1^2P_2=0, 
\end{split}
\label{eq:DispEq} 
\end{equation}
where 
\begin{equation}
\begin{split}
& F=\frac{J'_m(\kappa_1 a)}{J_m(\kappa_1 a)},~~~~~Q=\frac{H'^{(2)}_m(\kappa_2 a)}{H^{(2)}_m(\kappa_2 a)}, \\
& P_1 = \varepsilon_2\left[\frac{\varepsilon_1}{\varepsilon_2}\frac{J'_m(\kappa_1 a)}{J_m(\kappa_1 a)} - \frac{\kappa_1^2}{\kappa_2^2} \frac{H'^{(2)}_m(\kappa_2 a)}{H^{(2)}_m(\kappa_2 a)}\right], \\
& P_2 = \mu_2\left[\frac{\mu_1}{\mu_2}\frac{J'_m(\kappa_1 a)}{J_m(\kappa_1 a)} - \frac{\kappa_1^2}{\kappa_2^2} \frac{H'^{(2)}_m(\kappa_2 a)}{H^{(2)}_m(\kappa_2 a)}\right],
\end{split}
\label{eq:coeff} 
\end{equation}
and $\kappa_{1,2}^2=\omega^2 \varepsilon_{1,2} \mu_{1,2}-\beta^2$ are the transverse propagation constants inside and outside the nanowire, $J_m(\cdot)$ and $J'_m(\cdot)$ are the Bessel function of the first kind and its derivative with respect to the function argument, and $H^{(2)}_m(\cdot)$ and $H'^{(2)}_m(\cdot)$ are the Hankel function of the second kind and its derivative, respectively. 

A formal substitution of $m = 0$ into the dispersion equation (\ref{eq:DispEq}) related to the non-symmetric hybrid modes does not instantly lead to appearance of two separate dispersion equations related to the axially symmetric transverse magnetic (TM$_{0n}$; $\vec E = \{E_\rho,0,E_z\}$, $\vec H = \{0, H_\varphi, 0\}$) and transverse electric (TE$_{0n}$; $\vec E = \{0, E_\varphi, 0\}$, $\vec H = \{H_\rho, 0, H_z\}$) modes, as it is the case for ordinary dielectric waveguides. In general, in the waveguides coated by a resistive film the electric current ($\vec j = \sigma\vec E$) induced on the film by the electromagnetic field of corresponding mode can simultaneously possess both longitudinal $j_z$ and transverse $j_\varphi$ components. Such current appears also for the field having zeroth order ($m = 0$) resulting in the axially symmetric modes which nevertheless are hybrid ones. Therefore, in order to derive the separate dispersion equations related to the TM$_{0n}$ and TE$_{0n}$ modes the condition has to be imposed that the electric current possesses only a single component \cite{Veselov_book_1988}, i.e., either the condition $j_\varphi=0$ or $j_z=0$ should be satisfied yielding the corresponding dispersion equation
\begin{equation}
\frac{\varepsilon_1}{\kappa_1}\frac{J_1(\kappa_1 a)}{J_0(\kappa_1 a)} - \frac{\varepsilon_2}{\kappa_2} \frac{H^{(2)}_1(\kappa_2 a)}{H^{(2)}_0(\kappa_2 a)} = \frac{i \sigma}{\omega},
\label{eq:DispEqTM} 
\end{equation}
or
\begin{equation}
\frac{\kappa_1}{\mu_1}\frac{J_0(\kappa_1 a)}{J_1(\kappa_1 a)} - \frac{\kappa_2}{\mu_2} \frac{H^{(2)}_0(\kappa_2 a)}{H^{(2)}_1(\kappa_2 a)} = i\sigma\omega,
\label{eq:DispEqTE} 
\end{equation}
related to the TM$_{0n}$ modes and TE$_{0n}$ modes, respectively. In practice, similar to the case of cylindrical dielectric waveguides coated by a resistive film \cite{Veselov_book_1988}, the axially symmetric modes can be excited in a graphene-coated nanowire either by using specific excitation methods or by creating a particular anisotropy of the graphene sheet \cite{Kuzmin_NP_2018}.

By solving the corresponding dispersion equation  one can obtain the dependence of the complex propagation constant $\beta$ of the $m$-th order non-symmetric hybrid EH$_{mn}$ and HE$_{mn}$ modes or axially symmetric TM$_{0n}$ and TE$_{0n}$ modes on frequency or geometrical parameters of the waveguide. Since complex dispersion equations (\ref{eq:DispEq}), (\ref{eq:DispEqTM}) and (\ref{eq:DispEqTE}) cannot be solved analytically, the complex root search algorithm based on the M\"uller's method \cite{Mathews_book_1999} is adopted to calculate dispersion curves of $\beta$ numerically.

In our numerical calculations both Ohmic losses in the nanowire core and spatial dispersion of the graphene conductivity are neglected. Nevertheless, the model presented here is a good approximation to the real structures, since, as it was recently demonstrated in \cite{Lovat_2012, Fuscaldo_2015}, these factors have a negligible impact on the modal characteristics of the complex waves and thus can be ignored. 

\section{Dispersion Features of Complex Waves}
\label{sec:results}

Mathematically speaking, a graphene-coated nanowire is an open electromagnetic structure whose related boundary-value problem is described by a non-self-adjoint operator \cite{Veselov_book_1988}. The eigenvalues of the non-self-adjoint boundary-value problem in the general case are known to be complex values. As a result, the main feature of the given structure is a mandatory presence of complex waves in its spectra, i.e., the waves propagating through the graphene-coated nanowire under study are characterized by the complex propagation constants even though the energy dissipation in the media is absent.

\begin{figure*}[!ht]
\centering
\includegraphics[width=0.9\linewidth]{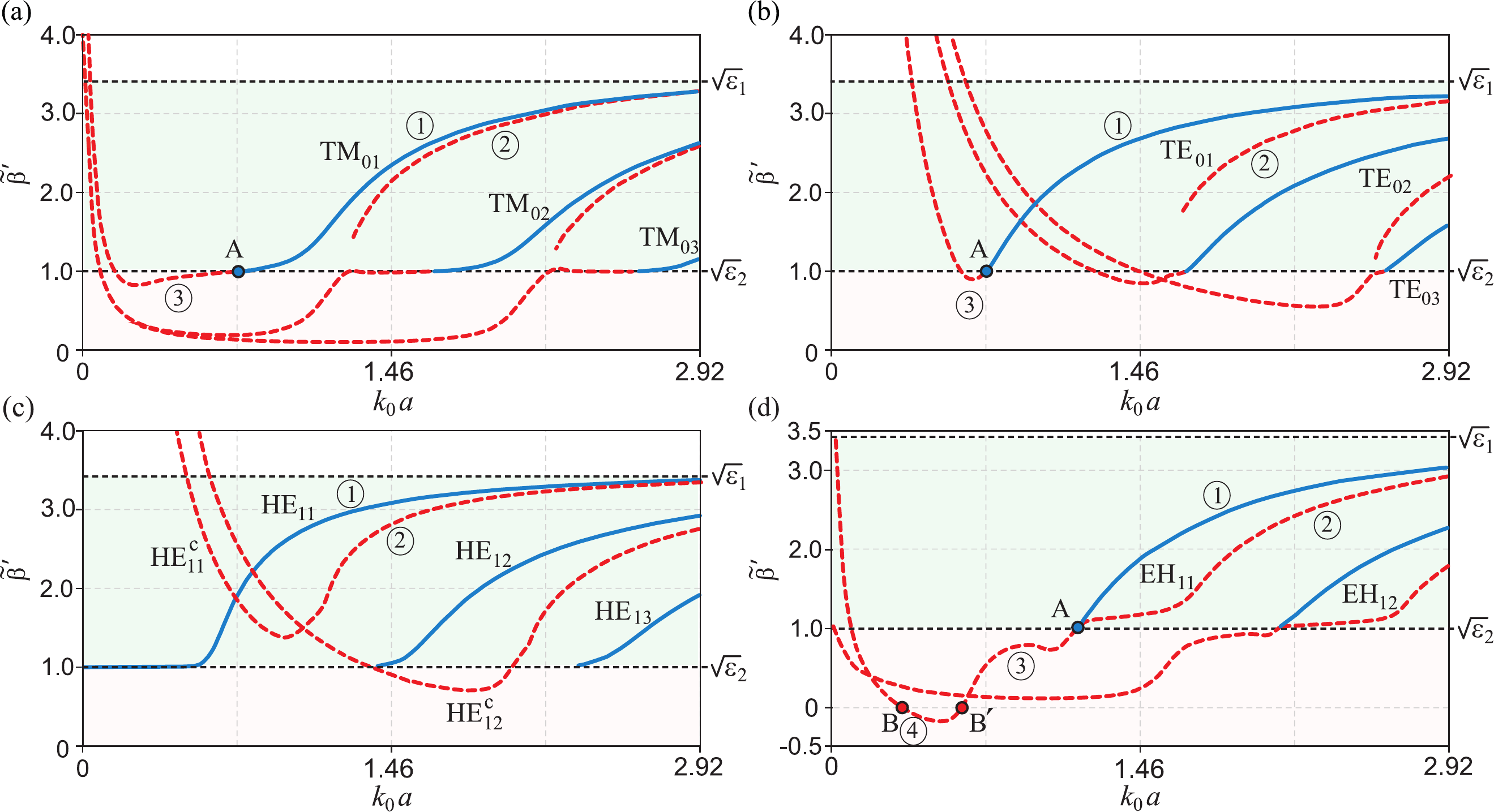}
\caption{Dispersion curves of axially symmetric (a) TM$_{0n}$ and (b) TE$_{0n}$ modes, and non-symmetric hybrid (c) HE$_{1n}$ and (d) EH$_{1n}$ modes ($n = 1,2,3$). Fragments of curves colored in blue (solid lines) and denoted by symbol \textcircled1 correspond to the trapped surface waves (guided modes), whereas those colored in red (dashed lines) and denoted by symbols \textcircled2, \textcircled3, and \textcircled4 correspond to the leaky waves (radiative modes). The points denoted by letter A mark the cut-off frequencies, whereas points denoted by letters B and B' mark the frequencies of transition between different types of the leaky waves. We set: $\varepsilon_1=\varepsilon_0 \varepsilon_{\textrm{Si}}(\omega,T)$ , $\varepsilon_2=\varepsilon_0$, $\mu_1=\mu_2=\mu_0$, $a=500$~nm, $T=300$~K, and $\mu_c=0.5$~eV.} 
\label{fig:fig_2}
\end{figure*}

The complex waves can possess different nature being evanescent, guiding, radiative or surface ones. Thus, the severe problem appears as to the methodology of the waveguide modes distinction and classification based on the traditional approaches, which always deal with purely real propagation constants and therefore can lead to some violent interpretations. Further in this paper a consistent method of the complex waves classification is involved \cite{Tamir_IEEE_1963, Ishimaru_book_1991} which is based on analysis of characteristics of the phase and attenuation constants derived from the complex longitudinal and transverse propagation constants. According to the method, the waves are initially attributed to the group of either `proper' or `improper' waves, wherein their type is then determined (see Appendix~A).

In order to classify the modes, the characteristics of the waves outside the waveguide ($\rho>a$) should be considered. In general, both longitudinal and transverse propagation constants are complex quantities which can be denoted as follows: $\beta = \beta' + i\beta'' = \beta' - i\alpha$ and $\kappa_2 = \kappa' + i\kappa'' = \kappa' - i\alpha_t$, where $\beta'$, $\alpha$ and $\kappa'$, $\alpha_t$ mean the longitudinal and transverse phase and attenuation constants, respectively. The waves propagating along the waveguide in the $+z$ direction ($\beta'>0$) can be attributed to the corresponding group considering the condition $\exp(-\alpha_t\rho)$ regarding their magnitude when $\rho \to +\infty$. If $\alpha_t >0$, the wave's magnitude decays exponentially in the transverse direction as $\rho$ increases, therefore, the wave is attributed to the `proper' waves. Otherwise, if $\alpha_t <0$, the wave's magnitude increases exponentially in the transverse direction as $\rho$ increases, and the wave is attributed to the `improper' waves, since it (physically) violates the radiation condition (nevertheless, such a wave exists in real systems provided that although its magnitude increases in the transverse direction, it decays exponentially in the longitudinal direction; e.g., they are in the basis of leaky-waves antennas \cite{Correas-Serrano_IEEE_2015}). It is obvious that depending on the signs of $\beta'$, $\alpha$, $\kappa'$, and $\alpha_t$, a variety of wave types can be distinguished (see Ref.~\cite{Ishimaru_book_1991} and Tab.~A.1). Further we are interested in three particular types of waves, namely trapped surface waves, leaky waves, and surface plasmons (these cases are denoted by letters C, H, and D in Tab.~A.1, respectively). 

\subsection{Trapped Surface and Leaky Waves}
\begin{figure*}[!htb]
\centering
\includegraphics[width= 0.9\linewidth]{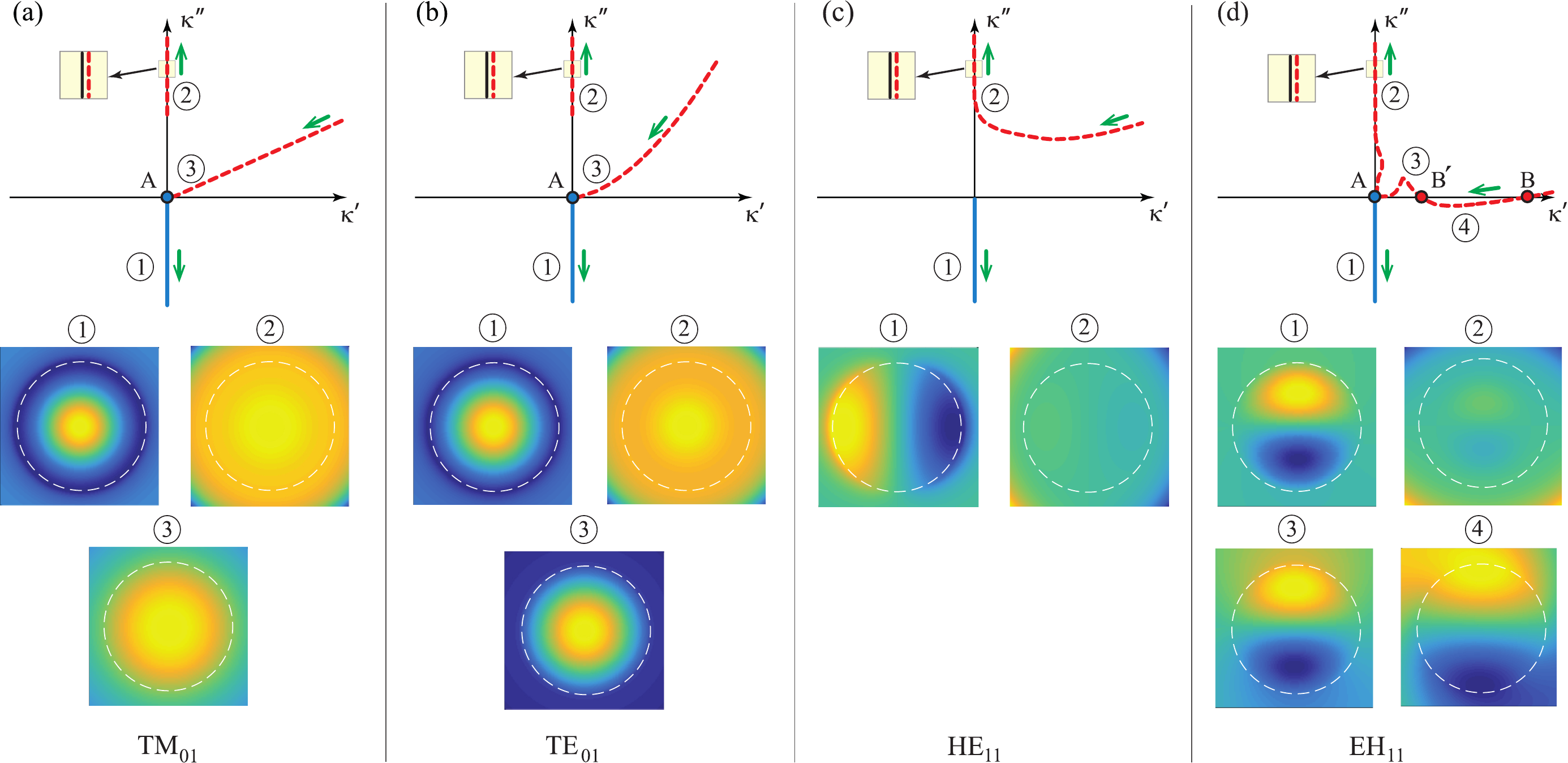}
\caption{Roots of dispersion equations in the complex $\kappa'-\kappa''$ plane, and the electric and magnetic field patterns of (a)~TM$_{01}$ ($E_z$), (b)~TE$_{01}$ ($H_z$), (c)~HE$_{11}$ ($H_z$), and (d)~EH$_{11}$ ($E_z$) modes. The half-planes $\kappa'' > 0$ and $\kappa'' < 0$ correspond to the regions of existence of proper and improper waves, respectively. Bold green arrows are directed toward the frequency increasing. Symbols \textcircled1-\textcircled4 and letters A, B and B$^\prime$ correspond to those depicted in Fig.~\ref{fig:fig_2}. In the electric field pattern  plots the white dashed circles designate the interface between the graphene-coated nanowire and free space. All problem's parameters are the same as in Fig.~\ref{fig:fig_2}.}
\label{fig:fig_3}
\end{figure*}
For the slow waves whose phase velocity $v_p$ is lower than the velocity of light $c$ (i.e., $\beta' = \omega/v_p > k_0 = \omega/c$), under the condition $\alpha=\kappa'=0$ (case C in Tab.~\ref{tab:CWtypes}) the waves carry a finite amount of power along the surface without attenuation and they decay exponentially in the transverse direction when $\rho \to +\infty$. Because of the attenuation due to $\alpha_t$, these waves are mostly concentrated closely to the interface, thus they are said to be `trapped' near the surface \cite{Ishimaru_book_1991}. In the modes' classification such trapped surface waves are attributed to the proper waves and, thereby, they belong to the discrete spectrum associated with the open waveguide system. Under the set of conditions $\alpha<0$, $\kappa'>0$, and $\alpha_t>0$ the waves are attributed to the proper leaky waves (case B in Tab.~\ref{tab:CWtypes}).

For the fast complex waves ($\beta' < k_0$), their magnitude decays exponentially in the longitudinal direction ($\alpha > 0$), whereas it increases exponentially in the transverse (radial) direction ($\alpha_t < 0$). Such waves are attributed to the improper waves and called leaky waves, since their energy is perennially leaked out from the interface (case H in Tab.~\ref{tab:CWtypes}). It is noteworthy that a particular solution of dispersion equations (\ref{eq:DispEq}), (\ref{eq:DispEqTM}) and (\ref{eq:DispEqTE}) can be related either to the trapped surface wave or to the leaky wave considering different branches of the dispersion curve of a particular waveguide mode. 

In order to demonstrate this feature clearly the dispersion curves of the lowest orders ($n =1,2,3$) axially symmetric TM$_{0n}$ and TE$_{0n}$ modes and non-symmetric hybrid EH$_{1n}$ and HE$_{1n}$ modes of the graphene-coated semiconductor nanowire under study are calculated and plotted in Fig.~\ref{fig:fig_2} on the frequency scale normalized on the nanowire radius. Here $\tilde{\beta}' = \beta'/k_0$ is an effective mode index, and curves colored in blue (solid lines) correspond to the proper trapped surface waves (guided modes), whereas those colored in red (dashed lines) correspond to the proper and improper leaky waves (radiative modes). Additionally, the roots of dispersion equations disposed on the complex $\kappa'-\kappa''$ plane as well as the electric and magnetic field patterns (magnitudes of the $E_z$ and $H_z$ components for the TM$_{01}$, EH$_{11}$ modes and TE$_{01}$, HE$_{11}$ modes, respectively) are presented in Fig.~\ref{fig:fig_3} for the modes with $n=1$. 

One can see that the region of existence of the trapped surface waves (branches \textcircled1) lies in the range $\sqrt{\varepsilon_2} \leq \tilde{\beta}' \leq \sqrt{\varepsilon_1}$, which is typical for the open dielectric waveguides. The dispersion curves of all modes start out at the line where $\tilde{\beta}' = 1$ and they approach the line $\tilde{\beta}'=\sqrt{\varepsilon_1}$ asymptotically. The mode cut-offs are at the points where $\kappa_2 = 0$ (points A), except the HE$_{11}$ mode which does not have any cut-off (direct expressions for the mode cut-offs of the trapped surface waves one can find in \cite{Marcuse_book_1991, Unger_book_1977}). 

In addition to the above discussed typical dispersion curves of the trapped surface waves inherent to the ordinary open dielectric waveguides, the presence of a graphene sheet coating the nanowire leads to the appearance of an additional set of branches in the dispersion curves lying in the range $\sqrt{\varepsilon_2} \leq \tilde{\beta}' \leq \sqrt{\varepsilon_1}$ for both axially symmetric TM$_{0n}$, TE$_{0n}$ and non-symmetric hybrid EH$_{1n}$ and HE$_{1n}$ modes (branches \textcircled2; they are also zoomed in and distinguished in Fig.~\ref{fig:fig_3} by the black arrows). 
\begin{figure}[!tb]
\centering
\includegraphics[width= 0.9\linewidth]{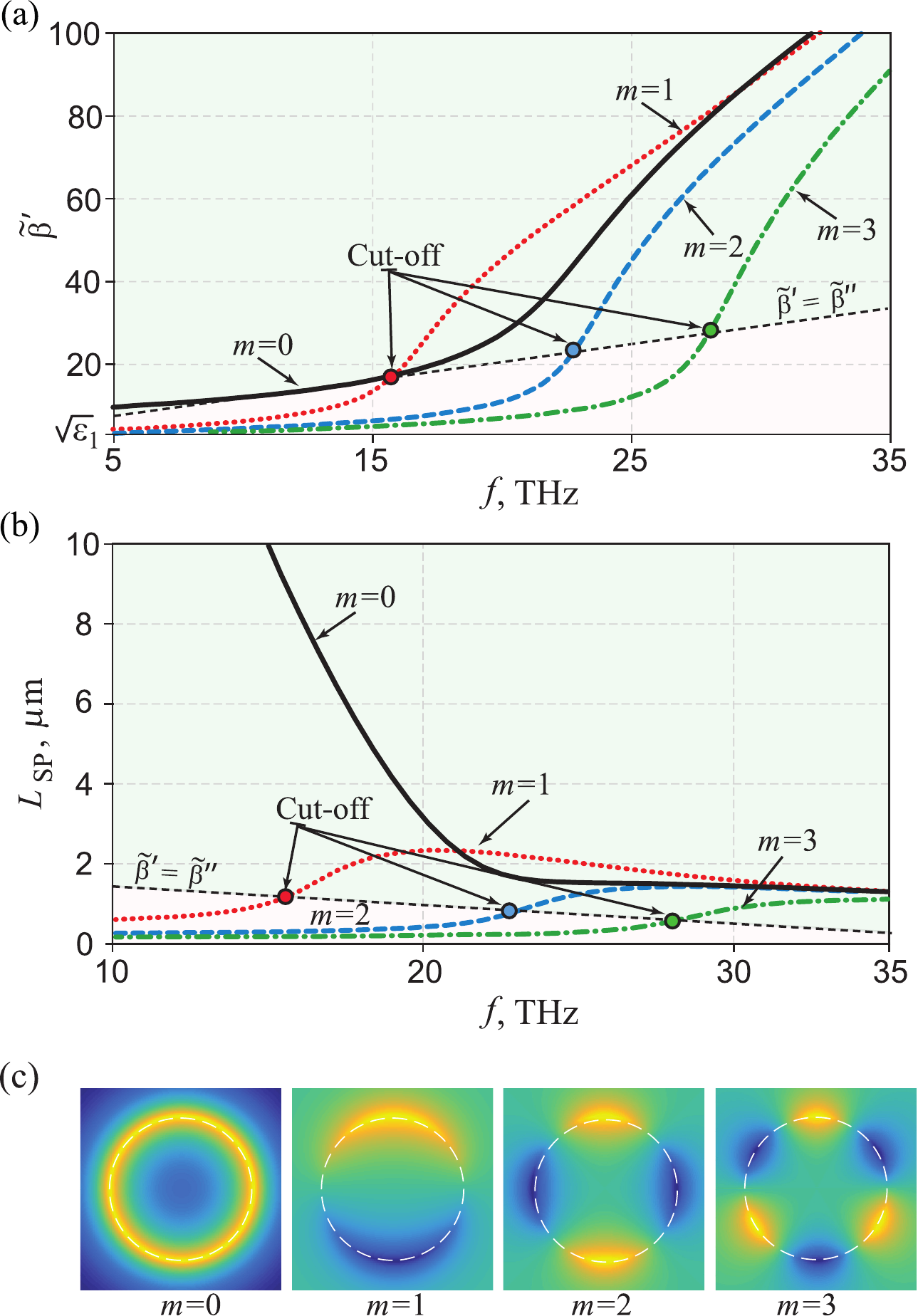}
\caption{(a) Dispersion curves, (b) propagation length, and (c) electric field patterns (magnitude of the $E_z$ component) of the TM modes of surface plasmons in the graphene-coated nanowire for the different azimuthal mode index $m$. The areas filled in light green correspond to the regions of propagating modes; $a = 50$~nm; patterns are plotted at the frequency $f=30$~THz.}
\label{fig:fig_4}
\end{figure}
\begin{figure}[!tb]
\centering
\includegraphics[width= 0.9\linewidth]{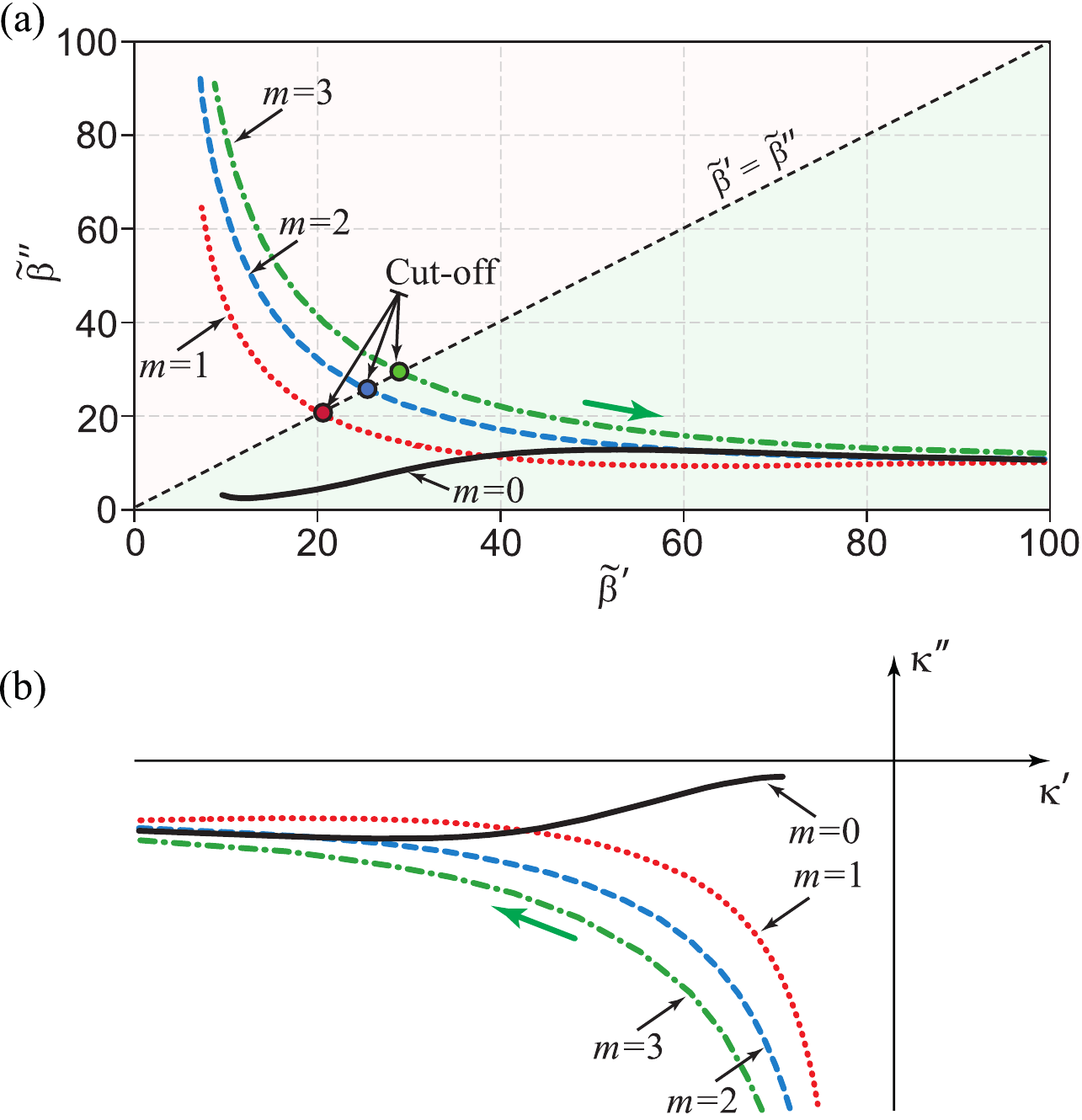}
\caption{Roots of the dispersion equation of the TM modes of surface plasmons in the graphene-coated nanowire plotted in the complex (a) $\tilde{\beta}'-\tilde{\beta}''$ and (b) $\kappa'-\kappa''$ planes. Bold green arrows are directed toward the frequency increasing. The area filled in light green corresponds to the region of propagating modes; $a = 50$~nm.}
\label{fig:fig_5}
\end{figure}
Considering characteristics of the phase and attenuation constants derived from the corresponding roots of dispersion equations these waves are attributed to the improper leaky waves. They are slow waves since their root branches are located in the immediate vicinity of the imaginary axis in the complex $\kappa'-\kappa''$ plane. These curves are closely spaced with those of the proper trapped surface waves approaching and merging with them as the frequency increases. They tend to the same asymptotic line  $\tilde{\beta}' = \sqrt{\varepsilon_1}$ as it is shown in Fig.~\ref{fig:fig_2}.  

Furthermore, the complex waves possessing either exponentially increasing or decreasing magnitude in the transverse direction can be found below the cut-off frequencies of the trapped surface waves. They are fast and slow waves, respectively \cite{Arnbak_1969, Veselov_book_1988}. Complex waves of the first type are an extension of the trapped surface waves (branches \textcircled3) and they arise from their cut-off points. At the same time, roots of dispersion equations corresponding to waves of the second type are far from cut-offs. 

In particular, regarding the axially symmetric TM$_{0n}$ and TE$_{0n}$ modes, dispersion curves of the trapped surface waves transform initially to those of the fast improper leaky waves which then turn into curves of the slow improper leaky waves as the frequency decreases. These curves demonstrate a rapid growth ($\tilde{\beta}' \to \infty$) when $\omega \to 0$. Moreover, the region of existence of the fast leaky waves has a tendency to expand as the index $n$ increases, and this region is wider for the TM$_{0n}$ modes compared to that of the TE$_{0n}$ modes for the same index $n$.  

These transitions between the different wave types are also observed for the dispersion curves of the non-symmetric hybrid HE$_{1n}$ and EH$_{1n}$ modes. However, the transitions inherent in the EH$_{11}$ mode have some peculiarities. Its dispersion curve has a very distinctive branch located between the points B and B$^\prime$ (branch \textcircled4) where the field magnitude in the transverse direction becomes to be decreasing exponentially with a distance far from the nanowire, i.e., a proper leaky wave arises in the region. This branch characterizing a proper leaky wave can exist only when the nanowire permittivity $\varepsilon_1$ is high enough. It is revealed that in order to realize such branches for the higher order modes the permittivity of nanowire should be further increased.

Another distinctive feature of the non-symmetric hybrid HE$_{1n}$ modes is the absence of transformation of the trapped surface waves into the leaky waves below their cut-offs. Instead, for this type of modes there are waves (denoted here as HE$_{1n}^\text{c}$) which are always complex in the whole frequency band $\omega \in \left[0, \infty \right)$. They are attributed to the slow improper leaky waves.

\subsection{Surface Plasmons}

A main dispersion feature of a graphene-coated semiconductor nanowire is its ability to support propagation of surface plasmons above the plasma frequency of semiconductor forming the nanowire, i.e., in the macroscopic description in such a waveguide surface plasmons exist even in the case when permittivities of the core and surrounding medium have the same sign. In consideration of (\ref{eq:effective}) surface plasmons appear due to the presence of an intermediate thin layer having negative permittivity ($\varepsilon_g'<0$) which acts as a conducting interface. Remarkably, both TM and TE modes of surface plasmons can propagate on this interface. 

In general, surface plasmons are complex waves and they can exist on a graphene sheet only when the imaginary part of the graphene conductivity satisfies several specific conditions \cite{Mikhailov_PRL_2007}. The TM modes exist when $\sigma''>0$ which holds in the frequency band limited by inequality $\hslash\omega/\mu_c<1.667$. The TE modes can appear when $\sigma''<0$. This condition holds in the frequency window $1.667<\hslash\omega/\mu_c<2$, which includes frequency of interband transition. An additional condition for the existence of the TE modes of surface plasmons is the inequality $|\sigma'| \ll |\sigma''|$, which also is dependent on the value of the chemical potential $\mu_c$ \cite{Kuzmin_SR_2016}. Moreover, the most favorable condition for the TE modes of surface plasmons propagation is the equality of the core and surrounding medium permittivit $\varepsilon_1=\varepsilon_2$ \cite{Kuzmin_SR_2016}. For the graphene-coated semiconductor nanowire under study there are no TE modes satisfying the existence conditions in the frequency band of interest, therefore in the subsequent discussion only dispersion characteristics of the TM modes of surface plasmons are under consideration.  

The dispersion equations for the TM modes of surface plasmons are obtained from Eqs.~(\ref{eq:DispEq}) and (\ref{eq:DispEqTM}) providing the following substitutions: $\kappa_{1,2}^2=\beta^2 - \omega^2 \varepsilon_{1,2} \mu_{1,2}$; $J_m(\cdot) \to I_m(\cdot)$; $H_m^{(2)}(\cdot) \to K_m(\cdot)$, where $I_m(\cdot)$ and $K_m(\cdot)$ are modified Bessel functions of the first and second kinds, respectively. The results of our calculations are summarized in Figs.~\ref{fig:fig_4} and \ref{fig:fig_5} for the TM modes with azimuthal index $m = 0,1,2,3$. Deriving and analyzing the longitudinal and transverse propagation constants from the roots of dispersion equations, we can attribute these modes to the proper waves (case D in Tab.~\ref{tab:CWtypes}).

The TM mode with azimuthal index $m=0$ is cut-off free, whereas the higher order modes exist above their cut-offs. At a fixed frequency $f$ the number of TM modes of surface plasmons supported by the graphene-coated nanowire may be estimated using the function: 
\begin{equation}
N(f)=2\pi a~\text{Re}\left[ \frac{i f}{\sigma(f) c} \big(\varepsilon_1(f)+\varepsilon_2\big)\right].
\label{eq:Nf} 
\end{equation}
The condition $N(f)<1$ corresponds to the single mode regime, while relation $N = m$ gives the cut-off frequency of the corresponding $m$-th mode in the multimode regime. 

In \cite{Gao_OE_2014,Gao_OptLett_2014} it was stated that the cut-offs of higher order modes appear under the condition $\tilde{\beta}' < \sqrt{\varepsilon_1}$ providing their dispersion curves go down and terminate abruptly reaching some horizontal line where $\tilde{\beta'}=\sqrt{\varepsilon_1}$ (see Fig.~2c in \cite{Gao_OE_2014} and Fig.~2 in \cite{Gao_OptLett_2014}). Nevertheless, here it is revealed that these cut-offs do not belong to any horizontal line of constant $\tilde{\beta'}$, instead they are located on a sloping line where $\tilde{\beta}' = \tilde{\beta}''$. In the correct approach, the region of existence of surface plasmons should be divided into two sub-regions where the waves appear as propagating ($\tilde{\beta}'>\tilde{\beta}''$) or reactive ($\tilde{\beta}'<\tilde{\beta}''$)  ones. In fact, these sub-regions are separated by the line where $\tilde{\beta}' = \tilde{\beta}''$ as depicted in Figs.~\ref{fig:fig_4} and~\ref{fig:fig_5}a.
    
\section{Conclusion}
\label{sec:concl}
To conclude, we have investigated dispersion features of complex waves supported by a graphene-coated semiconductor nanowire operating in the terahertz frequency band ($10-250$~THz). All waveguide modes were attributed to two groups of either `proper' or `improper' waves based on characteristics of their phase and attenuation constants. Within these groups the waveguide modes are classified and recognized as the trapped surface waves, fast and slow leaky waves, and surface plasmons. The peculiarities of dispersion curves have been analyzed in details for the propagation constant of axially symmetric TM$_{0n}$ and TE$_{0n}$ modes, as well as non-symmetric hybrid HE$_{1n}$ and EH$_{1n}$ modes belonging to the trapped surface waves and leaky waves.   

The necessary conditions for propagation of surface plasmons on a graphene sheet covering the nanowire were discussed. Characteristics of the TM modes of surface plasmons, which are proper complex waves, were studied. The region of existence of surface plasmons was correctly identified and described by introducing two sub-regions where the surface plasmons appear as propagating and reactive waves. We have found that the cut-offs of higher order modes should be searched on the line separating these sub-regions, and the condition for this line was obtained.   
 
\section*{Appendix A. Classification of Complex Waves}
\label{sec:AppA}

Different combinations of signs of $\beta'$, $\alpha$, $\kappa'$, and $\alpha_t$ lead to different types of waves. These wave types are systematized and presented in \cite{Ishimaru_book_1991} in the forms of a table (Tab.~\ref{tab:CWtypes}) and a circle plotted on the complex $\kappa'-\kappa''$ plane (Fig.~\ref{fig:fig_A1}), where the waves whose transverse propagation constant $\kappa$ belongs to the lower half-plane are attributed to the proper waves, otherwise they are improper waves.
\renewcommand{\thetable}{A.\arabic{table}}
\begin{table}[htbp]
\centering
\caption{\bf Systematization of Proper and Improper Complex Waves}
\begin{tabular}{ccccccc}
\hline
 &Case & $\beta'$  & $\alpha$  & $\kappa'$  & $\alpha_t$ & \\
\hline
       &A& + & 0 & + & 0 & Fast (waveguide) mode\\
       &E& + & 0 & -- & 0 & Plane-wave incidence \\ 
\\
Proper &B& + & -- & + & + & Backward leaky wave\\
wave   &C& + & 0 & 0 & + & Trapped surface wave\\
       &D& + & + & -- & + & Surface plasmon\\
\\       
Improper &F& + & -- & --& --& \\
wave	 &G& + & 0 & 0 & -- & Untrapped surface wave\\
         &H& + & + & + & -- & Forward leaky wave\\
\hline
\end{tabular}
  \label{tab:CWtypes}
\end{table}
\renewcommand{\thefigure}{A.\arabic{figure}}
\setcounter{figure}{0}
\begin{figure}[htbp]
\centering
\includegraphics[width=0.5\linewidth]{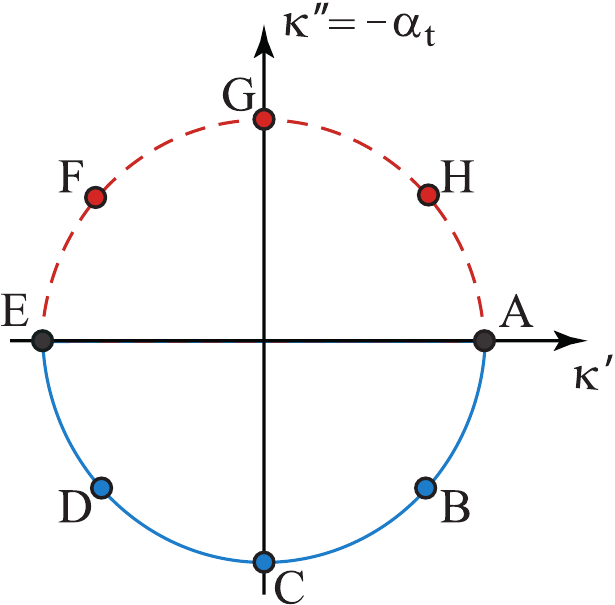}
\caption{Characteristic of proper (lower half-plane) and improper (upper half-plane) waves in the complex $\kappa'-\kappa''$ plane. Letters from A to H correspond to the appropriate cases depicted in Tab.~\ref{tab:CWtypes}}
\label{fig:fig_A1}
\end{figure}


\end{document}